\documentclass[twocolumn,trackchanges,twocolappendix]{aastex631}
\usepackage{natbib}
\usepackage{color}
\usepackage{graphicx}
\usepackage{multirow}

\renewcommand{\mathrm}[1]{#1}

\graphicspath{{./}}

\accepted{April 3, 2026}
\submitjournal{ApJL}

\shorttitle{First Calibration of the SBCR from O/B-type Stars in DEBs }
\shortauthors{Taormina et al.}

\begin{document}

\title{Toward Early-type Eclipsing Binaries as Extragalactic Milestones: First Calibration of the SBCR from O- and B-type Stars in Detached Eclipsing Binaries}

\correspondingauthor{M{\'o}nica Taormina}
\email{taormina@camk.edu.pl}

\author[0000-0002-1560-8620]{M{\'o}nica Taormina}
\affiliation{Centrum Astronomiczne im. Miko{\l}aja Kopernika PAN, Bartycka 18, 00-716 Warsaw, Poland}

\author[0000-0002-9443-4138]{G. Pietrzy{\'n}ski}
\affiliation{Centrum Astronomiczne im. Miko{\l}aja Kopernika PAN, Bartycka 18, 00-716 Warsaw, Poland}

\author[0000-0003-3861-8124]{B. Pilecki}
\affiliation{Centrum Astronomiczne im. Miko{\l}aja Kopernika PAN, Bartycka 18, 00-716 Warsaw, Poland}

\author{R.-P.  Kudritzki}
\affiliation{Institute for Astronomy, University of Hawaii at Manoa, Honolulu, HI 96822, USA}
\affiliation{LMU M\"unchen, Universit\"atssternwarte, Scheinerstr. 1, 81679 M\"unchen, Germany}

\author[0000-0002-7355-9775]{D. Graczyk}
\affiliation{Centrum Astronomiczne im.  Miko{\l}aja Kopernika PAN, Rabia{\'n}ska 8, 87-100 Toru{\'n}, Poland}

\author[0000-0002-0874-1669]{J. Puls}
\affiliation{LMU M\"unchen, Universit\"atssternwarte, Scheinerstr. 1, 81679 M\"unchen, Germany}

\author[0000-0002-3125-9088]{M. G\'{o}rski}
\affiliation{Centrum Astronomiczne im. Miko{\l}aja Kopernika PAN, Bartycka 18, 00-716 Warsaw, Poland}

\begin{abstract}
To measure precise distances beyond the Magellanic Clouds and determine an accurate value of the Hubble constant, eclipsing binary systems composed of early-type stars can play a crucial role. However, it is fundamental to first obtain a reliable empirical surface brightness-color relation (SBCR) for the hottest possible stars.
Based on our previous study of six detached eclipsing binaries composed of O- and B-type stars in the Large Magellanic Cloud, we calibrated the SBCR using 12 stars with $\mathrm{V-K_{\mathrm{s}}} < -0.6$ mag. 
We found a significant difference between O-type and B-type stars in SBCRs, which are clearly separated in mass. The relation based on B-type stars is consistent with the relation for redder stars from the literature. This allowed us to provide a combined relation valid for stars less massive than $\sim 16\,M_\odot$ in the wide color range $-0.9  <\mathrm{V-K_{\mathrm{s}}} < 2.1$ mag, with $\sigma = 0.025$ mag. Such a relation can provide extragalactic distances precise to as high as $\sim$1.2\% given the sufficient quality and number of target objects.
The relation for O-type stars ($\sigma = 0.055$ mag) remains uncertain due to its strong dependence on the method used to determine reddening and requires further study. However, we tested it on the only known eclipsing system in M33, and obtained distance modulus DM=$24.90 \pm 0.17 $ mag, which perfectly agrees with the published distance to the system.

\end{abstract}

\keywords{binaries: eclipsing --- stars: early-type --- stars: fundamental parameters --- Magellanic Clouds --- SBCR}

\section{Introduction} \label{sec:intro}

Distance determination in the Universe is crucial for understanding its structure and evolution. On a short scale, distances can be measured directly using geometric methods, but more often, empirical relations, theoretical models, or a combination of these methods are employed. However, the less model-dependent the technique, the more reliable it is.

Recently, the Hubble tension has become more evident. Still, the local measurements of the Hubble constant are primarily anchored to distances in the Milky Way and its close neighbor, the Large Magellanic Cloud (LMC), with one notable exception of NGC 4258 \citep{riess:2022, kudritzki:2024, freedman:2025}. It is thus vital to determine the most direct and least model-dependent distances to as many galaxies in the Local Group as possible.
For that reason, distances based solely on geometric methods, such as parallax, cannot be employed, and combinations of different aforementioned techniques must be considered.
Among them, the use of spectroscopically double-lined (SB2) detached eclipsing binaries (DEBs) as distance indicators is particularly important. A good example of this is the determination of the LMC distance based on B-type DEBs \citep{guinan:1998} and on late-type DEBs composed of giants in the LMC \citep{pietrzynski:2019} and the Small Magellanic Cloud \citep[SMC;][]{graczyk:2020}. For the late-type systems, a well-defined empirical surface brightness -- color relation (SBCR) was applied \citep{pietrzynski:2019}, but for B-type systems it was unavailable.
Generally, surface brightness (SB) relations link a light-emitting body's emerging flux per solid angle unit to its color or effective temperature \citep{kervella:2004}. If tight and well calibrated, these relations can yield accurate predictions of stellar angular diameters. 

Historically, the SBCR was constructed for stars in the solar neighborhood, whose angular diameters were measured with interferometry and lunar occultation. These diameters were then used to determine the stellar surface brightness $S_{\mathrm{V}}$ \citep[or $F_{\mathrm{V}}$, depending on the definition used,][]{wesselink:1969,barnes:1976a,barnes:1976b,barnes:1978}. Different color indices were used over time as variables in the relationship. \cite{diBenedetto:1998} showed that the scatter in the SBCR decreases for the color indices corresponding to a longer wavelength base, improving significantly when the color index $(\mathrm{V-K})$ is considered. For stars with $(\mathrm{V-K})> 0$ mag, the relationship allows a measurement of stellar angular diameters with a precision of 0.8\% \citep{pietrzynski:2019}, but for $(\mathrm{V-K})<0$ mag, the high scatter leads to an accuracy of 7.6\% in angular diameters \citep{challouf:2014}. These SBCRs were obtained in interferometric studies, but in the red region, a calibration based on eclipsing binaries is also available \citep{graczyk:2011}.

Here, we present the calibration of SBCR using DEBs with hot-star components ($T_{\mathrm{eff}} > 23000 \,K $, $(\mathrm{V-K_{\mathrm{s}}})_0 < -0.6$ mag for the first time.

DEBs composed of main-sequence late-type stars are very faint in more distant galaxies and cannot be observed with current instruments. Similarly, as in the Magellanic Clouds, one could look at systems composed of late-type giants. However, they are still relatively faint, have long orbital periods (of an order of a year or more), and exhibit narrow eclipses. This makes them very difficult to detect; in fact, none have been found yet. Moreover, collecting the necessary data for such systems would be very challenging.
If we want to measure geometrical distances to other galaxies in the Local Group farther than the LMC, we should use binary stars composed of main-sequence O- and B-type stars that are brighter and have periods as short as a few days. Thus, they can be readily detected and observed, particularly with the new generation of instruments and surveys.
However, to effectively use such early-type stars for distance determination, it is first necessary to advance in decreasing the scatter in the blue ($(\mathrm{V-K})_0 < 0$ mag) part of the SBCR.

In this context, we have observed and thoroughly analyzed a sample of double-lined detached eclipsing binaries composed of O- and B-type stars. This sample has been systematically studied across a series of papers: BLMC-01 and BLMC-02 (\citealt{taormina:2019}, hereafter Paper I); BLMC-02 (\citealt{taormina:2020}, Paper II); BLMC-03 (\citealt{taormina:2024}, Paper III); and BLMC-04, BLMC-05, and BLMC-06 (\citealt{taormina:2025}, Paper IV).

In this paper, we invert the method for determining distances to eclipsing binaries and use the known distances to the systems to determine the surface brightness ($S\mathrm{v}$) of their components. As all of them are located relatively close to the geometrical center of the LMC, individual distances to each system can be approximated with the precise distance to this galaxy \citep{pietrzynski:2019} corrected for the system position within the LMC. We use the LMC model of \citet{vandermarel2014} for this correction. Using the absolute radii of the stars from binary system modeling and the dereddened $V$ magnitudes of each component, we can calculate the surface brightness of each star.

\section{DEBs, the sample}  \label{sec:data}

In this series of papers, we have characterized six detached, double-lined eclipsing binaries in the LMC.
Our study demonstrated the diversity of configurations in which these systems can be found and revealed their complexity. Paper IV summarizes their properties.

The multiplicity of the systems is an important feature that can directly and indirectly affect the calibration and application of the SBCR. Extra light directly affects the observed brightness and, to a lesser extent, the color. If unaccounted for, orbital motion around the common center of mass may influence the entire binary model, including the absolute radii, further degrading the quality of the results. Therefore, proper multi-method detection and characterization of the system's multiplicity were crucial to our work.

After a detailed analysis and proper treatment of extra companions to the systems, we obtained accurate stellar parameters for 12 hot stars in the LMC, including their absolute radii. 
In Table~\ref{tab:systems}, we provide the basic properties of the systems, such as the orbital period and brightness of the eclipsing binary free of the light from additional companions (if any) in the $V$ band, as well as the components' radii and their spectral types. The latter are taken from the literature if available.

Because our methodology changed slightly after analyzing the first two systems, we reanalyzed BLMC-01 and BLMC-02 using the same method as applied to the remaining systems. Previously, we merged the same-band photometry from different surveys (OGLE, MACHO, EROS2) into a single light curve (LC), effectively having one LC per band. Moreover, $K_{\mathrm{s}}$ time series photometry (from VMC survey) was available only for the first system analyzed, and for BLMC-02, we used the 2MASS photometry corrected for our LC model.
Since Paper III, we have analyzed LCs from different surveys separately, thereby gaining greater control over the 3rd light in each band and survey. We also retrieved VMC data and obtained SOFI time-series photometry for all systems. In the current BLMC-02 modeling, we therefore included a $K_{\mathrm{s}}$-band light curve constructed from these two datasets.

After the reanalysis described above, the parameters of BLMC-02 remained essentially unchanged. For BLMC-01, the improved determination of the third light for each data set slightly changed the system inclination and the radius of the primary component (by about 1.7$\sigma$). In Table~\ref{tab:systems}, we provide the updated radii for BLMC-01 and 02, which are used for the SBCR calibration instead of those from Paper I.

\begin{deluxetable*}{lccccccc}
    \tabletypesize{\footnotesize}
	\tablecaption{Selected properties of the Early-type Eclipsing Binaries from our sample} \label{tab:systems}
         \tablehead{
                \colhead{Our ID} & \colhead{OGLE ID}  & \colhead{Orbital Period}  & \colhead{Brightness} & \colhead{$R_1$}   &  \colhead{$R_2$}   & Geom Correction\tablenotemark{*}  & \colhead{Spectral Type}  \\
                \colhead{  }     & \colhead{  }       & \colhead{[days]}         & \colhead{$V$[mag]}                   & \colhead{[$R_\odot$]  }   &  \colhead{[$R_\odot$]}  & \colhead{[mag]}   & \colhead{(from literature)} 
                }
           \startdata
                BLMC-01 & LMC-ECL-22270 &  5.4139927(1)  &  14.06   & 8.31 $\pm$ 0.09 & 12.98 $\pm$ 0.04 & -0.030 $\pm$ 0.008  &  B1III \tablenotemark{a}      \\
                BLMC-02 & LMC-ECL-06782 &  4.27077640(3)  & 13.72    & 8.83 $\pm$ 0.08 & 7.92 $\pm$ 0.10  & 0.032  $\pm$ 0.009  &  O7.5$+$O7.5 \tablenotemark{b}       \\
                BLMC-03 & LMC-ECL-21568 &  3.2254367(3)  & 14.34    & 7.70 $\pm$ 0.05 & 6.64 $\pm$ 0.06  & -0.026 $\pm$ 0.007  &  O9V$+$O9.5V \tablenotemark{c}       \\
                BLMC-04 & LMC-ECL-17660 &  6.2290957(4)  & 14.30    & 6.99 $\pm$ 0.03 & 7.61 $\pm$ 0.03  & -0.016 $\pm$ 0.005  &  ---\\
                BLMC-05 & LMC-ECL-18794 &  5.946812(2)  & 13.26    & 7.17 $\pm$ 0.07 & 14.21 $\pm$ 0.07 & -0.024 $\pm$ 0.007  &  O9.5III$+$B0  \tablenotemark{d}    \\
                BLMC-06 & LMC-ECL-05764 &  5.7259370(4)  & 14.21    & 9.08 $\pm$ 0.04 & 10.10 $\pm$ 0.04 & 0.019$\pm$0.006  &  B1.5III$+$B1.5III \tablenotemark{e}   \\           
	   \enddata
                \tablecomments{Observed $V$-band brightness of the system with the subtracted third light is provided. References: a) \cite{muraveva:2014}, b) \cite{taormina:2020}, c) \cite{massey:2012}, d) \cite{evans:2015}, e) \cite{guinan:1998}}
                \tablenotetext{*}{Following \cite{vandermarel2014}}
\end{deluxetable*}

\subsection{IR photometric data}
Currently, the most commonly used color index for the SBCR is $(\mathrm{V-K_{\mathrm{s}}})$, which we also use in this work.
The near-infrared $K_{\mathrm{s}}$ photometric data we employ come from two sources. The first is the VISTA Magellanic Clouds survey \citep[VMC,][]{cioni:2011}, which provides, on average, 15 measurements per system; the second is our observations collected with the SOFI imaging camera on the 3.58m ESO NTT \citep{moorwood:1998} at La Silla Observatory, Chile. Collecting additional data was crucial for improving light-curve coverage, particularly during eclipses, which is indispensable for proper characterization and subtraction of the third light. In total, we have 25-33 $K_{\mathrm{s}}$-band measurements per object. 

The $K_{\mathrm{s}}$ magnitude for each component of the binary systems results from the simultaneous modeling of light and radial velocity (RV) curves. The procedure is the same as for other bands, including the $V$ band, which is essential to avoid additional systematic errors (note that due to their lower availability, infrared data are often treated differently in the literature). As a result, we can subtract the measured third light from the observed brightness.
Moreover, if companions to a binary were identified, we corrected the light curve for the modulation of the signal arrival time induced by them (the so-called light travel-time effect).
For more details about the collection, reduction, and usage of IR data, please refer to previous papers of this series (in particular Papers III and IV).

For one of the systems, BLMC-04, although we have 26 photometric measurements in the $K_{\mathrm{s}}$ band, only two fall during eclipses, and neither is near their centers. In such a case, the result depends strongly on the point nearest to the center, with any measurement error propagating strongly to the third light. Indeed, although for other systems with third light the contribution in $K_{\mathrm{s}}$ is similar to that in other bands, for BLMC-04 it is about twice as high, while no clear wavelength dependence is seen from Gaia $G_{\mathrm{BP}}$ to Cousins $I$.
Specifically, from fitting the $K_{\mathrm{s}}$-band light curve, we obtained a third light of 10\%. In contrast, in other bands it varies from 3\% to 6\%, depending primarily on the survey rather than on wavelength. Between the bluer and redder bands of the same survey, differences are lower than 0.01, with varying signs, indicating colors near zero mag. In light of this, for BLMC-04, we adopted $0.05 \pm 0.05$ as the third light contribution.

\subsection{Reddening: Maps vs Interstellar Sodium Line}

In addition to the third light, reddening is another important factor in this study. As explained, e.g., in \cite{pietrzynski:2019}, in the red part of the SBCR ($\mathrm{V-K} \geq 0$ mag), the reddening vector is parallel to the relation and does not significantly affect the measurements. However, this is not the case for stars of spectral types earlier than A0, for which the relation gets steeper and is not linear. 

The reddening toward an object of interest is typically determined from reddening maps, which can be based on different traces. Examples of the most popular reddening maps of the LMC include those based on red clump (RC) stars \citep{haschke:2011,gorski:2020,skowron:2021}. 
These three maps show differences in the data used, the adopted zero color of the RC (reddening values depend strongly on it), and the method used to estimate the central value of $(\mathrm{V-I})_0$, among others. More importantly, these types of maps provide only spatially averaged values and, in general, are based on objects of a different kind from those we want to study.

Therefore, the most optimal approach is to determine the reddening directly to the object. One such method uses the calibration
between the equivalent width of the interstellar absorption sodium doublet, \ion{Na}{1}~D$_1$ at  5890.0~\AA{} and the color excess $\mathrm{E(B-V)}$ \citep{munari:1997}. This relation is calibrated for Galactic O- and B-type stars, which do not exhibit sodium in their spectra, and for which the broad stellar lines are easily distinguished from the narrow interstellar lines. Some deviations from this relation may be expected for the LMC. However, on average, more than half of the total reddening is due to the Milky Way, and differences in the total value should be small.

Papers I and III provide an example of the application of this method, with plots showing the observed spectra and a fit to multiple \ion{Na}{1}~D$_1$ components. 
The total $\mathrm{E(B-V)}$ values calculated from this sodium doublet for the six DEB systems analyzed until now in this series of papers are listed in Table~\ref{tab:reddComp}. These values were used to deredden our photometric data, using the reddening law of \citep{cardelli:1989} with $R_{\mathrm{V}} =3.2$ (the same as used in the calibration of \citealt{munari:1997}). For comparison, we also provide $\mathrm{E(B-V)}$ values from two recent LMC reddening maps in that table.

To calculate $\mathrm{E(B-V)}$ using data from \citeauthor{gorski:2020} we use the following formula: 

$$\mathrm{E(B-V)} = ( (\mathrm{V-I}) - (\mathrm{V-I})_{\mathrm{0,RC}} ) / 1.318\,,$$

\noindent where $(\mathrm{V-I})$ is the observed color of the RC and $(\mathrm{V-I})_{0, RC}$ is its intrinsic color. In their paper, several values are provided for the latter, depending on the calibration method used \citep[Table 5 in ][]{gorski:2020}. Here, we adopted $(\mathrm{V-I})_{\mathrm{0,RC}}=0.854$ mag based on \ion{Na}{1}~D$_1$ measurements of 20 late-type giant binaries in the LMC. 
\citeauthor{skowron:2021}, in their reddening map, provide the mean and peak values of the color excess in the $V$ and $I$ bands, $\mathrm{E(V-I)_{\mathrm{RC}}}$. Here, we considered the mean value and converted it to $\mathrm{E(B-V)}$ using the conversion factor 1.318, as above. 
In both cases, the considered field size around each binary is 3.4 arcmin (this is the only field size available in \citeauthor{skowron:2021}). In general, the reddening measurements follow the same trend but vary significantly across methods. Notably, our measurements for B-type stars are consistent within errors with those of \citeauthor{skowron:2021}. The highest difference is observed for the O-type system BLMC-05.

\begin{deluxetable}{lccc}
    \tabletypesize{\footnotesize}
	\tablecaption{Measured and estimated $\mathrm{E(B-V)}$ values} \label{tab:reddComp}
        \tablehead{
            \colhead{DEB ID} & \colhead{\ion{Na}{1}~D$_1$}     & \colhead{Górski+(2020)}   & {Skowron+(2021)}  
                 }
           \startdata
                BLMC-01  &  $0.193\pm0.017$    &  $0.26\pm0.11$   &  $0.22\pm0.17$  \\ 
                BLMC-02  &  $0.070\pm0.010$    &  $0.08\pm0.05$   &  $0.05\pm0.04$  \\ 
                BLMC-03  &  $0.143\pm0.013$    &  $0.19\pm0.10$   &  $0.22\pm0.17$  \\
                BLMC-04  &  $0.080\pm0.009$    &  $0.12\pm0.06$   &  $0.09\pm0.06$  \\
                BLMC-05  &  $0.052\pm0.005$    &  $0.18\pm0.09$   &  $0.16\pm0.13$  \\
                BLMC-06  &  $0.072\pm0.007$    &  $0.10\pm0.05$   &  $0.07\pm0.06$          
            \enddata
\tablecomments{For \ion{Na}{1}~D$_1$ this is a measured value directly to the system. The other two are estimates from the reddening maps. In all cases, the unit is a magnitude.}
\end{deluxetable}

\section{Surface Brightness -- Color relation}
With all necessary values available, we can calculate the SB for individual components of our systems. For that, we will use the following SB definition \citep[$S_{\mathrm{V}}$,][]{wesselink:1969}:

$$ S_{\mathrm{V}} = 5 \, \log \,\theta + m_{\mathrm{v_0}}$$

\noindent where $\theta$ is the angular diameter of the star and $m_{\mathrm{v_0}}$ their dereddened observed $V$ magnitude.  
Using the nominal values for the solar radius and parsec\footnote{following the IAU 2015 Resolution B3 \citep{prsa2016}}, the angular diameter $\theta$ can be expressed as  

$$ \theta= 9.301 * R / d $$

\noindent where $R$ is the absolute radius of a star and $d$ is its distance. 

If we know the distance to an eclipsing binary system, these equations provide directly the value of $S_{\mathrm{V}}$ for its components.  This is the case for our objects, which are all in the LMC. The distance to this galaxy was determined with a very high accuracy of 1.1\% by \citet{pietrzynski:2019}: $ d = 49.59 \pm 0.09_{\mathrm{stat}} \pm 0.54_{\mathrm{syst}}\, \textrm{kpc}$

Because the LMC disk is slightly inclined, we will correct the individual distances to the systems for their position in the LMC using its geometry determined by \citet{vandermarel2014}. The applied geometric corrections are given in Table~\ref{tab:systems}.

Using the SB and color values for 12 individual stars (components of the six systems), we constructed a surface brightness-color diagram for O- and B-type stars and performed a linear fit to the data. The color range covered by our sample is $-0.9 < (\mathrm{V-K_{\mathrm{s}}})_0 < -0.6$ mag, making it the bluest ever considered for SBCR calibration.  
The linear relation is described as follows:

$$ S_{\mathrm{V}} = a (\mathrm{V-K_{\mathrm{s}}}) + b$$

Examining the positions of the 12 components of early-type systems in the SBCR diagram and the fit residuals, we observed two groups that appear to follow distinct relationships. The six components of O-type systems are significantly separated and lie systematically above those of B-type systems.

\begin{figure*}
    \begin{center}
        \includegraphics[width=0.7\linewidth]{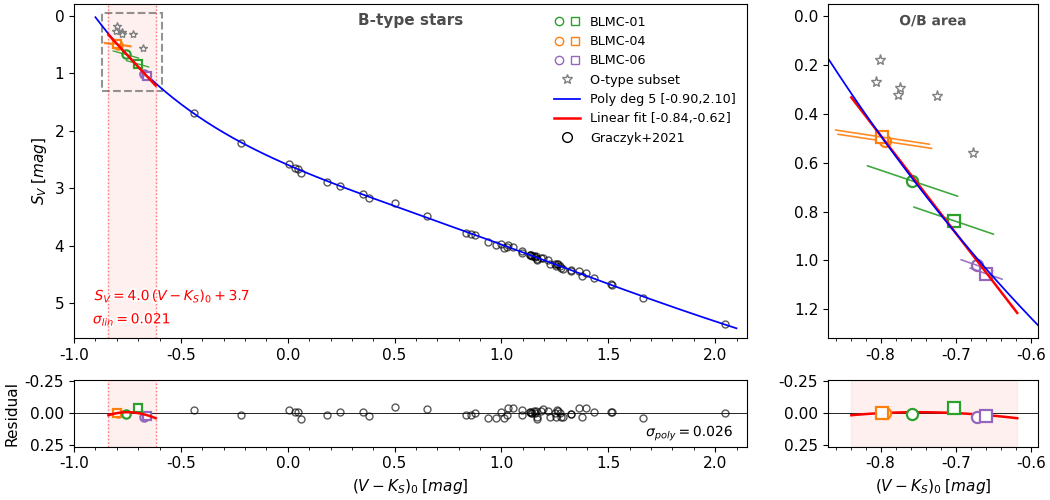} \\
        \vspace{0.4cm}
        \includegraphics[width=0.7\linewidth]{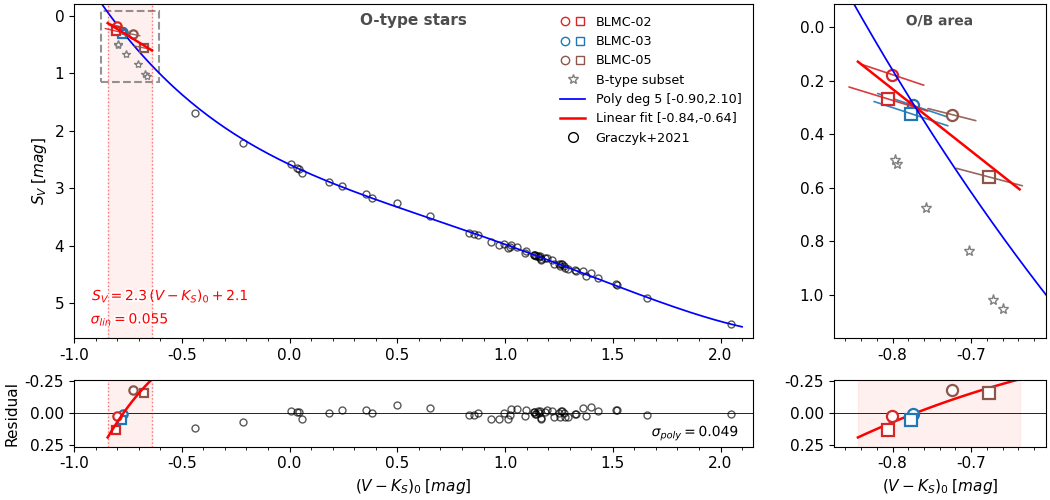} 
    \end{center}
\caption{Surface brightness $(\mathrm{V-K_{\mathrm{s}}})$-color relations considering either B (top panel) or O-type (bottom panel) stars from our sample. As the reddening is the main contributor to the uncertainties, the error bars are plotted along the reddening vector.
For polynomial fits, the relation is extended with redder DEBs (open circles) from \cite{graczyk:2021}. The linear fits to our sample in the color range [-0.9, -0.6] mag are shown as a red line. The fifth-order polynomial fits (to 62 components in both cases) are shown as blue lines, and the corresponding residuals are shown in the bottom panels. For B-type stars, the linear fit is tangent to the polynomial fit. For O-type stars, the polynomial fit yields significant systematic deviations. Our measured values of $S_{\mathrm{V}}$ and $(\mathrm{V-K_{\mathrm{s}}})_0$ are available in the electronic version.}
\label{sbcr_plot}
\end{figure*}

We thus divided our sample by spectral type into two groups and performed linear fits for each group separately. These fits are shown in  Fig.~\ref{sbcr_plot}, with the position of the other group for comparison. For B-type stars, the fitted parameters are: $a_B = 4.0$ and $b_B = 3.7$ ($\sigma_B=0.021$ mag), and for O-type stars, $a_O = 2.3$ and $b_O = 2.1$ ($\sigma_O=0.055$ mag).

To check for consistency and add an extra constraint, we included in our analysis 23 redder DEBs from \citet{graczyk:2021}. All except the bluest one (at $(\mathrm{V-K_{\mathrm{s}}})_0 \sim -0.44$ mag) were used by these authors for the calibration of the SBCR in the range [-0.2, 2.1] ($\sigma= 0.024$ mag).

With such an extended sample, we obtained the SBCR in the range [-0.9, 2.1] mag, fitting a fifth-order polynomial in the form of:

$$ S_{\mathrm{V}} = \sum_{i=0}^5 c_i (\mathrm{V-K_{\mathrm{s}}})^i$$

The polynomial fit using the B-type stars provides a tight relation with no systematic deviations in residuals (Fig.~\ref{sbcr_plot}; top panel). Moreover, the linear fit is nearly tangent to the polynomial one. This shows that B-type stars are consistent with the SBCR provided by \cite{graczyk:2021} for redder stars. Our data, however, extends the relation to much bluer ($\mathrm{V-K} \sim -0.9$ mag) objects.

The fitted coefficients are: $c_{0,B} = 2.593,\, c_{1,B} = 1.623,\, c_{2,B} = -0.622,\, c_{3,B} = 0.587,\, c_{4,B} = -0.236 $, and $c_{5,B} =0.032 $ ($\sigma_B=0.025$ mag). When using only O-type stars from our sample in the polynomial fit, we obtain the following coefficients, $c_{0,O} = 2.587,\, c_{1,O} = 1.790,\, c_{2,O} = -0.917 , c_{3,O} =0.672 , c_{4,O} =-0.155 $, and $c_{5,O} = -0.003 $ ($\sigma_O=0.049$ mag). 

In this case, the residuals exhibit greater scatter and systematic deviations, which means that O-type stars are not consistent with the red part of the SBCR. 
The linear fit is also less steep than the polynomial one in the corresponding range (Fig.~\ref{sbcr_plot}, bottom panel). Interestingly, what we would expect is a steeper slope for this type, because of a very low dependence of $(\mathrm{V-K})$ color on temperature, as it samples a deep part of the Rayleigh-Jeans tail.

Although a larger sample of B- and O-type stars would be needed to confirm whether the difference we see is general, it is clear within our sample. However, the reason for the observed difference in SBCR and its form remains unclear. This can be either an intrinsic feature that depends on factors such as mass or a difference in the surrounding environment that, for instance, affects the estimation of reddening. It is also unclear if this change is gradual or abrupt.

\begin{figure*}
    \begin{center}
        \includegraphics[width=0.9\textwidth]{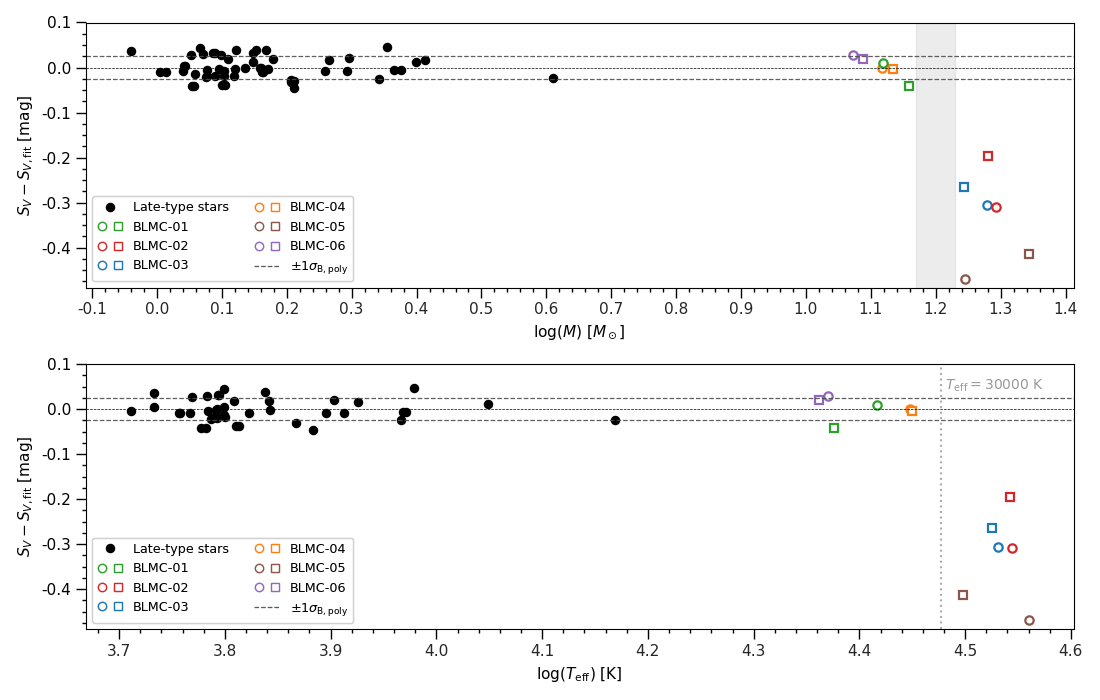}
    \end{center}
\caption{Surface brightness residuals from the polynomial SBCR based on B-type stars versus stellar mass (top) and effective temperature (bottom). Our B and O-type stars are plotted together with those of later types from \citet{graczyk:2021}. O-type stars clearly deviate, although their $(V-K_{\mathrm{s}})_0$ colors overlap significantly with those of B-type, and minimal extrapolation is needed. 
The shaded area marks a possible transition phase in mass between the B and O-type stars. The effective temperature of $30000\,K$ also clearly separates the deviating stars.
}
\label{sbres_mass}
\end{figure*}

We conducted several additional tests to shed light on these issues. In the top panel of Fig~\ref{sbres_mass}, we show the residuals from the polynomial fit to the extended B-type sample, including the O-type stars not used in the fitting, versus the stellar mass. The stellar masses of our O/B stars come from our binary models (see Papers I-IV). The masses of late-type stars from the literature were determined in a similar way, as all are components of eclipsing systems. The separation in mass is clear, although the scatter for the massive O-type stars is higher than for the rest of the sample. This test does not indicate whether the change is gradual, but the transition between the two groups must occur between $15\,M_\odot$ and $17\,M_\odot$ (shaded area in Fig~\ref{sbres_mass}, top panel). A similar result is obtained when the effective temperature, rather than mass, is used, with a transition zone around $30000\,K$ (Fig~\ref{sbres_mass}, bottom panel). 

\begin{figure*}
    \begin{center}
        \includegraphics[width=0.48\textwidth]{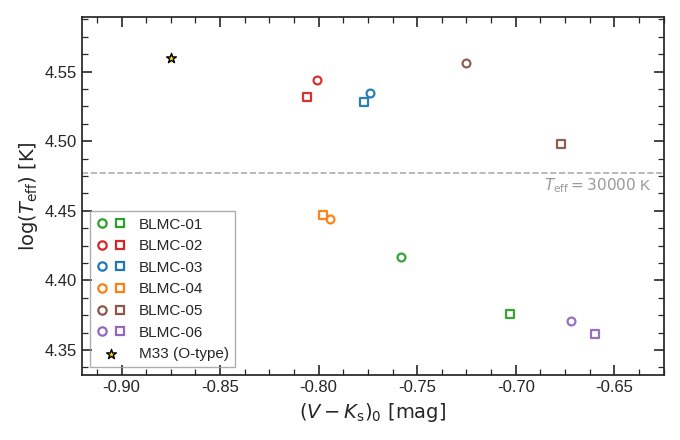}
        \includegraphics[width=0.48\textwidth]{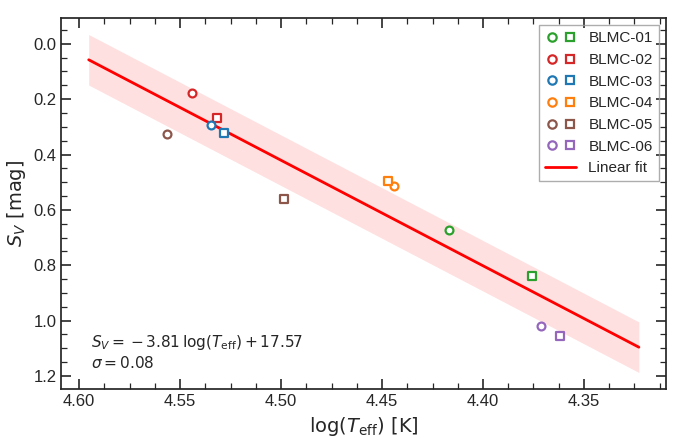}
    \end{center}
\caption{(left) Color -- temperature diagram for components of our systems. The colors clearly overlap, yet the temperatures of more massive O-type stars differ significantly from those of B-type ones. The position of the M33 O-type system is consistent with our O-type sample.  (right) Surface-brightness versus temperature. Although the scatter is higher, a common relation for O- and B-type stars can be derived.}
\label{sb_temp}
\end{figure*}

In another test, we compared the dereddened $(\mathrm{V-K_{\mathrm{s}}})_0$ colors with the derived temperatures (Fig~\ref{sb_temp}, left panel). As shown, stars with similar colors but different spectral types (or masses) have significantly different temperatures. As noted by \cite{pecaut:2013}, there is a long-standing problem with empirical color calibration for O-B2 stars \citep[][]{johnson:1953} due to their lack in the Local Bubble. The attempt of \citeauthor{pecaut:2013} to calibrate this type of stars resulted in the $(\mathrm{V-K_{\mathrm{s}}})$ color monotonically decreasing (shifting toward blue) with the increasing temperature. Still, to deredden the photometry, they used the Q-method \citep{johnson:1953,gutierrez:1975} which depends on the photometric data and the observed colors. On the contrary, our estimate of reddening based on the interstellar sodium doublet is entirely independent of photometry.

We also tested the relationship between the determined temperature and the surface brightness. In contrast to the SBCR, here the result is as expected: both values are clearly correlated, although the scatter is higher than for the $(\mathrm{V-K_{\mathrm{s}}})$ color (Fig.~\ref{sb_temp}, right panel). The last two tests suggest that for stars hotter than $\sim 30000K$, the color may not be a good indicator of surface brightness, as already shown by \cite{hummer:1988}. For such stars, the temperature determined through detailed spectroscopic analysis is much more reliable.

For our O-type stars, we also determined the reddening from SED fitting (see Paper II and Paper III). The derived $\mathrm{E(B-V)}$ values for BLMC-02, 03, and 05 are: 0.09 mag, 0.16 mag, and 0.09 mag with corresponding $R_{\mathrm{V}}$ values of 2.9, 3.2, and 3.0, respectively. Using these measurements, we obtain bluer colors, which move the O-stars close to the extension of the B-star relation. The reddening for B-type stars is still determined from the sodium doublet, so the comparison is not fully meaningful; however, it indicates that the sodium line method for O-type stars may miss some dust components (e.g., containing ionized Na) along the line of sight to the object. On the other hand, using the reddening derived from the SED fitting contradicts our goal of obtaining an empirical SBCR.

Finally, we test the SBCRs constructed based on the reddening maps, using the values listed in Table~\ref{tab:reddComp}. Applying the reddening from \citeauthor{skowron:2021}, we obtain results that are somewhat similar to our main results. B-type stars are also consistent with the redder stars, but both the linear and polynomial fits yield larger scatter, $\sigma=0.061$ mag and $\sigma=0.031$ mag, respectively. Four O-type stars appear too blue and two too red, hindering the use of them to extend the polynomial relation further. 
Using the reddening from \citeauthor{gorski:2020}, both B- and O-type stars are not consistent with redder stars and exhibit higher scatter. 
In conclusion, the use of statistically derived reddenings (not in the line of sight to the object) may be risky due to excessive random errors and systematics. However, for B-type stars, reddenings from the sodium line and the map of \citeauthor{skowron:2021} are consistent.

Currently, we recommend using the linear or polynomial fits obtained from components of B-type systems; however, their application to O-type stars remains unclear. If the behavior of these stars in the SBCR is an intrinsic feature, the corresponding relations based on O-type stars should be used. Alternatively, there may be an external factor that (at least for our sample) is correlated with spectral type and makes O-type stars appear redder. If so, using a correct relation may depend on the environment in which the given star is found.

\section{Calibration test - distance to M33}
Having new relations at hand, it is tempting to apply them to an eclipsing binary in a galaxy more distant than the Magellanic Cloud. Fortunately, one such system was studied previously by \cite{bonanos:2006}.
Using a different, more model-dependent method, these authors published the first distance determination to an O-type DEB in M33, called D33 J013346.2+304439.9. 
From the LC and RVC analysis, \citeauthor{bonanos:2006} obtained absolute masses and radii of the components in this system. Using $BVRJHK_{\mathrm{s}}$ photometry for the flux calibration, they determined the distance modulus (DM) to the system of 24.92 $\pm$ 0.12 mag (964 $\pm$ 54 kpc). This distance is longer than other determinations of M33 distance based, i.e., on Cepheids \cite[DM=24.62 $\pm$ 0.07 mag,][]{gieren:2013}. 

In \cite{bonanos:2006}, only system magnitudes are provided and not those of individual components of the M33 DEB. Therefore, in our calculations, we used a geometric average of the components' radii and their average surface brightness. To maintain homogeneity, we deredden the system color using the same method as for our systems, i.e., measuring the interstellar sodium line.
This procedure yielded $(\mathrm{V-K_{\mathrm{s}}})_0$ = -0.875 mag. We note that this is slightly below the range for our linear fits; however, we are primarily interested in comparing these relations, and the extrapolation is similar for both. The difference between the linear and polynomial fits for B-type stars at this color is only 0.04 mag, which is less than the other involved uncertainties.

Applying the linear calibration for only O-type stars from the sample, the distance to the M33 DEB is DM=24.90 $\pm$ 0.17 mag, consistent with the one found by \cite{bonanos:2006}.

For comparison, from the polynomial relation based on B-type stars we obtain DM=24.81 $\pm$ 0.33 mag. The higher uncertainty here arises from a steeper dependence on color, despite a significantly lower standard deviation than for the linear O-type relation.
Both distance moduli are highly correlated because their primary source of uncertainty is the same: measurement errors for the system in M33, which account for more than 88\% of the total uncertainty. 

It is interesting that the position of the M33 system naturally extends the color-temperature relation for our O-type sample (Fig.~\ref{sb_temp}, left), strengthening the possibility that something may affect the colors of stars above about $16\, M_\odot$ (or $30000\, K$).

\section{Discussion and Conclusions}\label{sec:summary}
For the first time, we constructed and fitted the SBCR using O- and B-type stars, which are components of detached eclipsing binary systems. The latter condition is essential because stars in such systems (i.e., with a similar rotation and oblateness) will be used to determine distances to other galaxies.

When using the full sample of O- and B-type stars, the scatter in the SBCR is approximately 0.08 mag, comparable to that reported in previous interferometric studies \citep{challouf:2014}. However, our more precise analysis, based on a larger sample of stars, revealed a significant systematic difference between B and O spectral types.

After complementing our sample with redder DEBs, we obtained a significantly better fit for the B-type components than for the O-type components.
The SBCR for B and later types reaches a scatter of only 0.021 mag for the linear fit and 0.025 mag for the polynomial fit. Such a relation can provide extragalactic distances with a precision of $\sim$1.2\% if high-quality data on DEBs in other galaxies are available. Due to the high consistency between B and later types, we recommend using this relation for effective temperatures $T_{\mathrm{eff}}<30000\,K$ and masses below ${\sim}16 \,M_\odot$, in the corresponding color range.  The polynomial fit is well defined for colors $(V-K_{\mathrm{s}})\in$ [-0.9, 2.1] mag, and the linear fit for [-0.84, -0.62] mag.

Regarding O-type stars, further study is needed to understand their observed deviation from the SBCR based on other spectral types. One possibility is an additional reddening from their neighborhood, which is not correlated with the presence of \ion{Na}{1} in the interstellar medium\footnote{Note, however, that the relation between the presence of interstellar sodium lines and reddening was calibrated on O- and B-type stars.}. In that case, however, we would need an additional reddening of approximately 0.04 mag for all three O-type systems, which, on average, amounts to almost 50\% of the measured values and is three times their estimated uncertainties.  Moreover, to obtain consistency, the additional reddening should be lower for the bluest, more massive O-type stars, and higher for the reddest, less massive ones. That is not consistent with the assumption that the effect appears for O-type stars and is not present for B-type stars.

Another issue could be the value of the total-to-selective extinction ratio. In our analysis, we assumed $R_{\mathrm{V}}$=3.2, as it was used in the calibration of the \ion{Na}{1} D method, while this value can commonly vary between 2 and 5 \citep{hur:2015,urbaneja:2017}. This could potentially affect the colors of stars. However, the $R_{\mathrm{V}}$ values determined from our SED fitting are consistent with the assumed value, ranging from 2.9 to 3.2. Therefore, we do not consider it as a possible reason for the deviation of our O-type stars.

Currently, our results suggest that it is not possible to obtain an empirical SBCR that is universally valid for all types of stars, because above an effective temperature of $\sim 30000\,K$, the color may not be a good indicator of surface brightness (Fig.~\ref{sb_temp}, left). However, the use of a separate relation for stars with $T_{\mathrm{eff}}>30000\,K$ is still disputable.
The results vary too much depending on the reddening source used.
To address this and the other problems mentioned above, we plan to analyze additional O/B systems in the LMC to confirm the observed difference and investigate its cause. 

Regarding the distance to the M33, it is more tempting to adopt the distance from the polynomial relation for B-type stars as the valid one. Still, the system in M33 is of O type, even earlier than the O-type LMC stars used in our calibration (see Fig.~\ref{sb_temp}, left). The use of both, however, yields very similar distances, with the first only slightly shorter. Comparison of these results with other distance determinations to M33 (not based on the same system) is hindered by the high uncertainty associated with the M33 binary, even though the relations themselves provide much higher precision. A solution is either to obtain higher-quality data for the system or to identify and analyze more similar systems in that galaxy.

\begin{acknowledgments}
The research leading to these results has received funding from the European Research Council (ERC) under the European Union's Horizon 2020 research and innovation program, grant agreement No. 951549 (project UniverScale). BP acknowledges support from the Polish National Science Centre grant SONATA BIS 2020/38/E/ST9/00486. RPK has been supported by the Munich Excellence Cluster Origins, funded by the Deutsche Forschungsgemeinschaft (DFG, German Research Foundation) under Germany's Excellence Strategy (EXC-2094, 390783311). We also acknowledge support from the Polish Ministry of Science and Higher Education grant 2024/WK/02.
\end{acknowledgments}

\bibliographystyle{aasjournal}
\bibliography{bibliogStars}

\end{document}